\documentclass{aastex61}

\shorttitle{The Mg~{\sc ii} line in lobe-dominated quasars}
\shortauthors{Wildy et al.}

\usepackage{longtable}

\begin{document}

\title{Mg II line properties in lobe-dominated quasars}

\correspondingauthor{Conor Wildy}
\email{wildy@cft.edu.pl}

\author{Conor Wildy}
\author{Bozena Czerny}
\author{Agnieszka Ku\'zmicz}

\affiliation{Center for Theoretical Physics, Polish Academy of Sciences \\
Al. Lotnik\'ow 32/46, 02-668 Warsaw, Poland}



\begin{abstract}

We investigate the relationships between Mg~{\sc ii} ${\rm \lambda{}}$2798 emission line properties, as well as between these properties and inclination angle and Fe~{\sc ii} strength, in a lobe-dominated quasar sample. We find no correlation between Mg~{\sc ii} line width and inclination, unlike previous studies of the general quasar population. This suggests that the Mg~{\sc ii} emission region in these objects is not of a thin disk geometry, however the average equivalent width of the line negates a spherical alternative. A positive correlation between Mg~{\sc ii} equivalent width and inclination cannot be ruled out, meaning there is no strong evidence that Mg~{\sc ii} emission is anisotropic. Since thin disk emission would be highly directional, the geometric implications of these two findings are compatible. The lack of line width correlation with inclination may also indicate that Mg~{\sc ii} is useful for estimating black hole masses in lobe-dominated quasar samples, since it is unbiased by orientation. Some quasars in our sample have almost edge-on viewing angles and therefore cannot have a smooth toroidal obscurer co-planar with the accretion disk. Alternatives may be a distorted dusty disk or a clumpy obscurer. This could result from the sample selection bias towards high inclination objects, rather than intrinsic differences between lobe-dominated and typical quasars. Five objects have visible [O~{\sc iii}] allowing equivalent width calculation, revealing it to be higher than in typical quasars. Since these objects are of high inclination, this finding supports the positive correlation between [O~{\sc iii}] equivalent width and inclination found in a previous study.

\noindent

\end{abstract}

\keywords{galaxies: active -- quasars: emission lines -- quasars: general}



\section{Introduction}

Quasars are a luminous and distant sub-category of active galactic nuclei (AGN), which are powered by supermassive black holes (SMBHs) accreting matter at the centers of galaxies. The quasar ultraviolet (UV) and optical spectrum, consisting of emission lines superimposed on a power-law continuum together with varying degrees of absorption, may be used as an indicator of the structure close to the central engine. Several studies within the past few decades have investigated apparent correlations between spectral properties. A very notable example is the application of Principal Component Analysis to a quasar sample by \citet{boroson92}, which found that the primary eigenvector, \emph{Eigenvector 1} (EV1), indicated a strong anti-correlation between the emission line strengths of Fe~{\sc ii} and [O~{\sc iii}]. The secondary component, \emph{Eigenvector 2} (EV2), revealed an anti-correlation between optical luminosity and He~{\sc ii} strength. A popular explanation for the EV1 correlation is that it is due to the distribution of Eddington ratio (L/${\rm L_{Edd}}$) values, where L is the quasar luminosity and ${\rm L_{Edd}}$ is the Eddington luminosity corresponding to the mass of the black hole \citep{boroson02,shen14}. A possible mechanism for this could be the ``puffing up'' of the accretion disk at high L/${\rm L_{Edd}}$, resulting in a region relatively close to the black hole being ionized by thermal x-rays, thereby generating excess Fe~{\sc ii} \citep{boroson02}. 

An alternative (or addition) to the Eddington ratio-driven model is one based upon an orientation dependence of the observables. Orientation effects have long been used to explain differences in AGN categories. For example Type 1 AGN, which exhibit broad emission lines, are theorized to be observed at low inclinations with respect to the accretion disk axis, resulting in an unobscured view of the central engine, while Type 2 AGN, which do not show broad lines in unpolarized light, are viewed at high inclination angles. This results in obscuration by a dusty torus of those environs close to the SMBH from which the broad emission lines are thought to originate, known as the broad line region (BLR) \citep{urry95}. In \citet{risaliti11} the distribution of [O~{\sc iii}] equivalent widths (EWs) in quasar optical spectra obtained using the Sloan Digital Sky Survey (SDSS) was found to be consistent with a scenario in which [O~{\sc iii}] EW increases with increasing inclination angle (see \citet{adelman08} for more details on the SDSS). This was based on the assumption of a geometrically thin, optically thick accretion disk \citep{shakura73}, whose apparent optical emission decreases with increasing inclination angle, while [O~{\sc iii}] emits isotropically far from an obscuring torus \citep{mulchaey94} and has a luminosity proportional to the continuum bolometric luminosity. The distribution of broad line EWs (H${\rm \beta{}}$, Mg~{\sc ii} ${\rm \lambda{}}$2798 and C~{\sc iv} ${\rm \lambda{}}$1549) did not follow the pattern found by \citet{risaliti11} for [O~{\sc iii}], suggesting the sources of these lines radiate anisotropically. A possible explanation for this is broad line emission resulting from a flattened disk around the black hole \citep{vestergaard00,pozonunez14}.

The dependence of broad emission line width on [O~{\sc iii}] EW was investigated by \citet{bisogni17}. They found that the H${\rm \beta{}}$, H${\rm \alpha{}}$ and Mg~{\sc ii} lines in a large quasar sample showed a strong correlation between [O~{\sc iii}] EW and three different line width measurements. These were full-width at half maximum (FWHM), line dispersion and the inter-percentile velocity defined as spanning the wavelength ranges accounting for 90 per cent of the line flux. They interpreted this result as supporting a disk-like configuration for the BLR, assuming [O~{\sc iii}] EW is a strong indicator of inclination angle following \citet{risaliti11}. They invoked two further pieces of evidence to support this conclusion, first that the distribution of the [O~{\sc iii}]/H${\rm \beta{}}$ ratio shows a similar power-law tail at high values to that seen for [O~{\sc iii}] EW, and second that broad lines showing a double-peaked structure are more common at high than at low [{O~{\sc iii}] EWs.

Radio data has been used in several studies to infer the degree of quasar inclination. The work of \citet{orr82} discussed the ratio of quasar core to extended 5~GHz radio flux ($R$) and its relationship to orientation, $R$ being expected to be smaller in quasars with large inclination angles. This can be understood in the context of schemes which unify quasar radio properties by orientation. In such schemes, lobe-dominated quasars ($R$$<$1) require a significant projection of the radio-axis in the plane of the sky, whereas core-dominated ($R$$>$1) quasars have a comparatively axis-on viewing angle, leading to a higher fraction of the radio emission appearing from the core regions \citep{antonucci93,saikia99,allen00}. Later, the ratio of radio core luminosity to optical continuum luminosity, denoted $R_{v}$, was introduced by \citet{wills95} and presented as a better inclination measure than $R$. This was due to $R_{v}$ providing a tighter correlation than $R$ with two parameters in a sample of quasars, namely a derived inclination angle obtained by examining motion within the jet structure \citep{ghisellini93} and broad emission line FWHM. More recently, \citet{aars05} compared both $R$ and $R_{v}$, as well as a ``pseudoangle'' based on the rank orderings of $R$ and the projected linear size of the radio emission, originally described in \citet{hough02}, with several emission line widths. In all three cases they found correlations supporting larger Mg~{\sc ii} FWHM at higher inclination angles.

This paper investigates, for a sample of SDSS quasars, the interdependencies between various Mg~{\sc ii} ${\rm \lambda{}}$2798 line properties, as well as correlations between these properties and two other characteristics, namely quasar inclination angle and Fe~{\sc ii} emission strength. The inclinations of each quasar in our sample were obtained from a previous study (\citet{kuzmicz12}, hereafter Paper~I), within which radio data enabled calculation of the angles. Reverberation mapping \citep{blandford82} reveals that lower ionization lines such as Mg~{\sc ii} have longer time delays between continuum and line variation than more highly ionized lines \citep{fine13}, implying a stratified BLR structure where the ionization parameter decreases with increased radial distance from the black hole. Therefore, a study of Mg~{\sc ii} line properties potentially reveals information about the BLR ``layer'' encompassing low ionization conditions. A preliminary version of the results was presented during the conference "Quasars at all Cosmic Epochs " held in Padova, Italy in April 2017, and published in \citet{wildy17}. Here we present a full description of the methodology, we also expand the study
by discussing more correlations and adding information from the IR band. We also report the results in much greater detail. 

The contents of this paper are organized as follows: Section~2 describes the sample selection of quasar spectra obtained from the SDSS, Section~3 discusses the line and continuum fitting process, Section~4 details the line properties and correlation analysis, Section~5 discusses the implications of the results in Section~4 for the lobe-dominated quasar population, and Section~6 provides conclusions.

\section{Sample selection}

The quasars used in this study were obtained from the sample of 43 giant radio sources (GRSs) and 48 smaller comparison objects compiled in Paper~I having calculated inclination angles. These quasars were selected from published radio source data or from radio source catalogues (see references therein). Quasar radio properties were measured in Paper~I using maps from the National Radio Astronomy Observatory Very Large Array Sky Survey (NVSS) and the Faint Images of the Radio Sky at Twenty Centimeters (FIRST) survey (see \citet{condon98} and \citet{becker95} respectively). All quasars in their sample were double-lobed, being of Fanaroff Riley type II classification \citep{fanaroff74} and were required to have an observed radio structure spanning at least 0.2 arcmin in angular diameter, rendering the lobes and core resolvable in all data sources used. The GRSs are defined in Paper~I as those objects with a radio structure $>$0.72~Mpc, given the cosmological parameters listed in \citet{spergel03}. Their results indicated no significant differences in fundamental quasar properties (such as accretion rate and black hole mass) between the two populations, so no distinction between them is made in this study. Most GRSs in their sample had a redshift suitable for studying the Mg~{\sc ii} line using ground-based optical observations. Their comparison sample objects were selected explicitly to render the Mg~{\sc ii} line visible in optical spectra, hence redshifts for all objects are within the range 0.4$<$$z$$<$2.1. 

Calculation of the inclination angle ($\theta{}_{i}$) was performed in Paper I for all but one quasar in their sample. This was done using the flux from the two visible radio lobes of each object and assuming that Doppler boosting was the principal contributor to the asymmetries in the observed radio lobes, using the following equation:

\begin{equation}
\label{eqn:incang}
\theta{}_{i}=\rm{acos}\left(\frac{1}{\beta{}_{j}}\frac{\left(s-1\right)}{\left(s+1\right)}\right)\, ,
\end{equation}

\noindent with $s=(S_{j}/S_{cj})^{\frac{1}{\left(2-\alpha{}\right)}}$, $S_{j}$ and $S_{cj}$ being the peak flux density of the lobes appearing closer to and further from the core respectively, and $\beta{}_{j}$ being jet velocity as a fraction of the speed of light, which was assumed to be 0.6 in accordance with \citet{wardle97} and \citet{arshakian04}. The spectral index was fixed at $\alpha{}$=$-$0.6 as in \citet{wardle97}. Additionally, uncertainties in the inclination angles were estimated, using typical peak flux density measurement errors. It would be expected that the selection of lobe-dominated quasars with a large angular size would lead to the sample having a strong bias towards high inclination angles. This is indeed the case, the average of all $\theta{}_{i}$ values recorded in Paper~I being 73$^{\circ}$. Other biases in the sample are noted in Paper~I (see Section 3 therein), most notably the fact that double-lobed radio quasars are very rare within the population of SDSS quasars, comprising only 1.7 per cent of the total \citep{devries06}. Therefore results obtained here may not necessarily be applicable to the quasar population in general.

All but five of the 91 objects in the Paper~I sample have at least one SDSS spectrum. Some objects have repeated SDSS spectroscopic observations, for such objects we only used the spectrum designated ``sciencePrimary'' in Data Release 13. Before any analysis was performed, all SDSS spectra were corrected for Milky Way dust extinction using the method of \citet{cardelli89}, with $A_{V}$ values obtained from the NASA/IPAC Extragalactic Database (NED). The spectra were then corrected to the quasar rest-frame using redshift values ($z$) obtained from the NED and subsequently rebinned onto a 1~\AA{} grid.  We only analyzed those objects where the SDSS spectrum spans at least 2300--3300~\AA{} in the quasar rest-frame, eliminating 30 objects from the final analysis sample. This criterion enables observation of the spectral region occupied by the Mg~{\sc ii} line as well as sufficient wavelength coverage either side of the line to enable fitting of the \emph{small blue bump}, which results from Fe~{\sc ii} and hydrogen in the BLR. The 56 remaining quasars along with their redshifts, inclination angles and signal-to-noise ratios (S/N) at 2700~\AA{} are listed in Table~\ref{tab:quasars}. Of these, only 45 were used in the final analysis sample, as a satisfactory fit to the small blue bump emission was not possible for ten objects (see Section~3.1), while one additional quasar was excluded due to a substantial absorption feature overlapping the Mg~{\sc ii} emission line. The distribution of inclination angles in the final 45-quasar sample is shown in Fig.~\ref{fig:hist}.

\begin{center}
\begin{longtable}{l l l l}
\caption{List of quasars having measured inclination angles and sufficient spectral coverage.}
\endfirsthead
\textit{Continued from previous page}\\
\hline
\endhead
\hline
$\dagger{}$ in object rest-frame\\
$\ddagger{}$ excluded from final sample due to poor SBB fit\\
$\ast{}$ excluded from final sample due to absorption feature\\
\textit{continued on following page}
\endfoot
\hline
\endlastfoot
\hline
SDSS name&Redshift&Inclination angle&S/N at 2700~\AA{}$^\dagger{}$\\
 & &($^{\circ}$)& \\
\hline
SDSS J005115.11$-$090208.5&1.259&55$\pm{}5$&9.95\\
SDSS J020448.29$-$094409.5&1.003&81$\pm{}5$&12.7\\
SDSS J021008.48$+$011839.6&0.870&63$\pm{}5$&21.5\\
SDSS J024534.06$+$010813.7$^\ddagger{}$&1.529&85$\pm{}6$&22.8\\
SDSS J075034.40$+$654125.6&0.747&65$\pm{}5$&16.5\\
SDSS J075448.86$+$303355.1&0.796&87$\pm{}6$&21.6\\
SDSS J080906.22$+$291235.4&1.481&28$\pm{}7$&38.7\\
SDSS J081240.08$+$303109.4&1.313&71$\pm{}5$&14.4\\
SDSS J081409.22$+$323731.9&0.843&72$\pm{}5$&13.4\\
SDSS J081735.07$+$223717.7$^\ddagger{}$&0.982&71$\pm{}5$&36.3\\
SDSS J081941.12$+$054942.6&1.689&81$\pm{}5$&3.56\\
SDSS J082806.83$+$393540.2$^\ddagger{}$&0.763&79$\pm{}5$&15.0\\
SDSS J083906.94$+$192148.7$^\ddagger{}$&1.692&41$\pm{}6$&29.7\\
SDSS J084239.96$+$214710.3&1.181&85$\pm{}6$&12.8\\
SDSS J090207.20$+$570737.8&1.592&79$\pm{}5$&18.4\\
SDSS J090429.62$+$281932.7&1.122&45$\pm{}5$&28.2\\
SDSS J090649.98$+$083255.8&1.616&86$\pm{}6$&14.8\\
SDSS J091858.15$+$232555.4&0.690&81$\pm{}5$&23.2\\
SDSS J092425.02$+$354712.6&1.344&84$\pm{}6$&23.0\\
SDSS J092507.27$+$144425.6$^\ddagger{}$&0.896&84$\pm{}6$&20.0\\
SDSS J094418.85$+$233119.9&0.989&83$\pm{}6$&19.8\\
SDSS J095206.38$+$235245.2&0.971&89$\pm{}6$&33.7\\
SDSS J095934.49$+$121631.5&1.091&73$\pm{}5$&7.41\\
SDSS J100017.67$+$000523.6$^\ddagger{}$&0.905&85$\pm{}6$&11.5\\
SDSS J100445.75$+$222519.3&0.981&85$\pm{}6$&17.2\\
SDSS J100507.07$+$501929.8&2.016&76$\pm{}5$&15.7\\
SDSS J100607.70$+$323626.1&1.026&80$\pm{}5$&8.11\\
SDSS J100943.55$+$052953.8$^\ddagger{}$&0.942&78$\pm{}5$&42.4\\
SDSS J102026.87$+$044752.0&1.134&61$\pm{}5$&6.19\\
SDSS J102041.14$+$395811.2&0.830&58$\pm{}5$&23.2\\
SDSS J102313.61$+$635709.2$^\ddagger{}$&1.194&70$\pm{}5$&40.7\\
SDSS J103050.90$+$531028.7&1.197&77$\pm{}5$&20.4\\
SDSS J105403.26$+$415257.5&1.093&75$\pm{}5$&15.9\\
SDSS J105636.25$+$410041.2&1.781&87$\pm{}6$&4.57\\
SDSS J111023.84$+$032136.1&0.966&85$\pm{}6$&11.2\\
SDSS J111858.62$+$382852.2&0.747&55$\pm{}5$&14.0\\
SDSS J111903.28$+$385852.5&0.735&64$\pm{}5$&17.3\\
SDSS J115139.68$+$335541.4&0.851&32$\pm{}6$&20.6\\
SDSS J121701.37$+$101952.9&1.884&87$\pm{}6$&15.3\\
SDSS J122925.53$+$355532.1&0.828&57$\pm{}5$&9.91\\
SDSS J123604.51$+$103449.2&0.667&63$\pm{}5$&15.9\\
SDSS J125607.67$+$100853.6&0.824&69$\pm{}5$&11.2\\
SDSS J131946.20$+$514805.7$^\ddagger{}$&1.055&61$\pm{}5$&40.5\\
SDSS J132106.65$+$374153.4&1.135&79$\pm{}5$&11.7\\
SDSS J133411.70$+$550124.9&1.247&89$\pm{}6$&17.1\\
SDSS J134034.70$+$423232.1&1.345&89$\pm{}6$&9.82\\
SDSS J135817.60$+$575204.5&1.372&85$\pm{}6$&38.8\\
SDSS J140806.20$+$305448.3$^\ast{}$&0.845&80$\pm{}5$&31.5\\
SDSS J143215.53$+$154822.3$^\ddagger{}$&1.005&87$\pm{}6$&15.9\\
SDSS J143334.31$+$320909.2&0.936&88$\pm{}6$&2.69\\
SDSS J151329.29$+$101105.5&1.548&83$\pm{}6$&36.0\\
SDSS J155002.00$+$365216.7&2.071&64$\pm{}5$&6.49\\
SDSS J155729.93$+$330446.9&0.944&89$\pm{}6$&17.2\\
SDSS J162229.93$+$353125.3&1.471&82$\pm{}5$&17.4\\
SDSS J162336.45$+$341946.3&1.994&13$\pm{}13$&6.98\\
SDSS J223458.73$-$022419.0&0.550&83$\pm{}6$&24.6
\label{tab:quasars}
\end{longtable}
\end{center}

\begin{figure}
\centering
\resizebox{\hsize}{!}{\includegraphics[width=8cm]{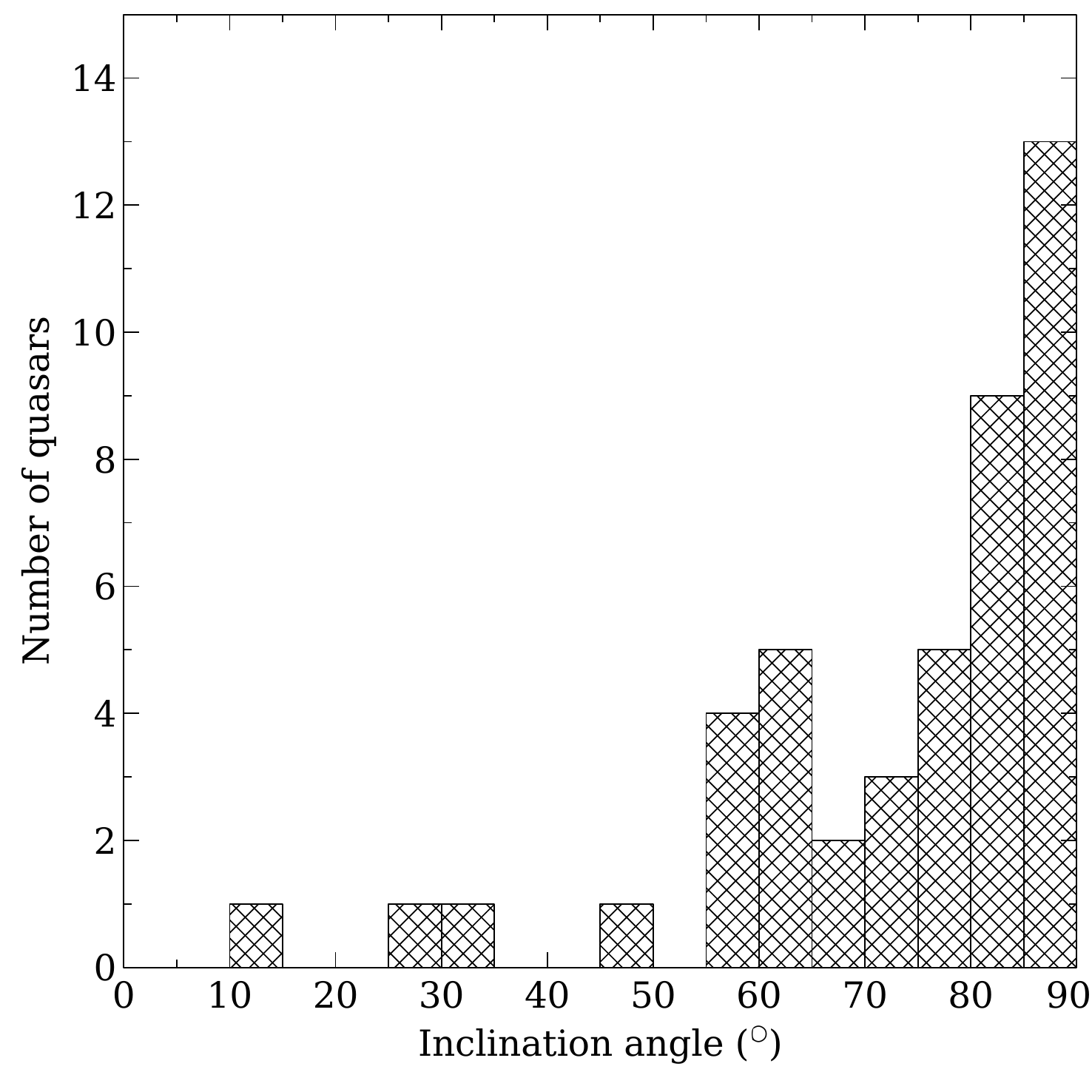}}
\caption{Distribution of inclination angles in the final quasar analysis sample, illustrated as a histogram.}
\label{fig:hist}
\end{figure}

\section{Spectral fitting}

\subsection{The small blue bump}

In addition to the disk power-law continuum, the Mg~{\sc ii} line sits on the small blue bump (SBB), a complex blend of emission features consisting of the Balmer continuum (BC) and many broadened and blended Fe~{\sc ii} lines \citep{jin12}. Therefore, the analysis of this line requires accurate modelling of the underlying SBB emission. Paper~I attempted to ``clean'' the spectral region near Mg~{\sc ii} by first fitting a power-law to several relatively-line-free disk-continuum bands outside the SBB in each quasar spectrum, before subtraction of this continuum. Subsequently, two Fe~{\sc ii} templates, one UV and one optical, were scaled and broadened to match the continuum-subtracted spectrum in the SBB region, within their respective wavelength domains, to form a total Fe~{\sc ii} model. Afterwards, this model was subtracted and the procedure repeated to ensure effective removal of the iron emission. The Fe~{\sc ii} templates used in Paper~I were those of \citet{vestergaard01} and \citet{veroncetty04}, based respectively on UV and optical data and obtained from observations of the narrow-line quasar I Zw 1.

Following on from \citet{wildy17}, we improve upon the Paper~I fitting procedure for our sample in two important ways. First, as an alternative to the two templates used in Paper~I, we utilize iron templates developed by \citet{bruhweiler08} to reconstruct the SBB Fe~{\sc ii} emission. These provide several model spectra produced for a range of physical conditions in the line generating plasma, the variable parameters being number density, microturbulence and ionizing flux. The best template could then be selected for any individual quasar in our sample based on the quality of the fit it provided. Second, we calculate the BC emission for each quasar based on the method described in \citet{grandi82} by using the following equation:  

\begin{equation}
\label{eqn:bccalc}
F_{\nu}^{\rm BC}=F_{\nu}^{\rm BE}e^{\left(-h-\nu{}_{\rm BE}\right)/\left(kT_{e}\right)}\, ,
\end{equation}

\noindent where $F_{\nu}^{\rm BC}$ is the flux of the BC at a given frequency ($\nu{}$), $F_{\nu}^{\rm BE}$ is the BC flux at the Balmer edge (located at 3646~\AA{}), $T_{e}$ is the electron temperature, $k$ is the Boltzmann constant and $h$ is the Planck constant. This continuum is ignored in Paper~I despite it being a significant contributor to the overall SBB feature. Initially, for each quasar spectrum in our sample, a power-law continuum was fitted to relatively line-free regions and an appropriate Fe~{\sc ii} template was selected by eye and broadened using Gaussian smoothing to approximately match the features of the SBB. Subsequently, a grid in iron template normalization, $T_{e}$ and $F_{\nu}^{\rm BE}$ was used to generate model SBB spectra. The gridpoint providing the model spectrum found to give the best fit (using chisquare minimization) to the power-law-subtracted observed spectrum in the range 2300--2600~\AA{}, 3000--3300~\AA{} was chosen as the appropriate SBB template for that particular quasar.

The iron template normalization values used for the grid were chosen to span 1--100 per cent, in 1 per cent increments, of the difference between the power-law continuum and the observed spectrum at 3000~\AA{}. For most objects, where the wavelength coverage included the Balmer edge, $F_{\nu}^{\rm BE}$ values spanned 1--100 per cent, in 1 per cent increments, of the difference between the power-law continuum and the observed spectrum at the Balmer edge wavelength. Otherwise, $F_{\nu}^{\rm BE}$ values were calculated based upon the same incremental process but at 3000 \AA{}, by extrapolating to the Balmer edge for each gridpoint based on Equation~\ref{eqn:bccalc}. The $T_{e}$ values spanned 5000--20\,000~K in increments of 1000~K. Finally the two components of the best-fitting model, together with the power-law continuum, were simultaneously re-scaled to the SBB region, maximizing the quality of the fit. This last step was performed using the \emph{specfit} software within the Image Reduction and Analysis Facility ({\sc iraf})\footnote{http://iraf.noao.edu/} \citep{iraf93}. An example of a high quality final fit to the SBB emission (excluding the Mg~{\sc ii} line, whose fitting is described in Section~3.2) is shown in Fig.~\ref{fig:sbbeg}. The fits were of poor quality in 10 out of the 56 cases, these quasars were therefore excluded from further analysis.

\begin{figure}
\centering
\resizebox{\hsize}{!}{\includegraphics[width=8cm]{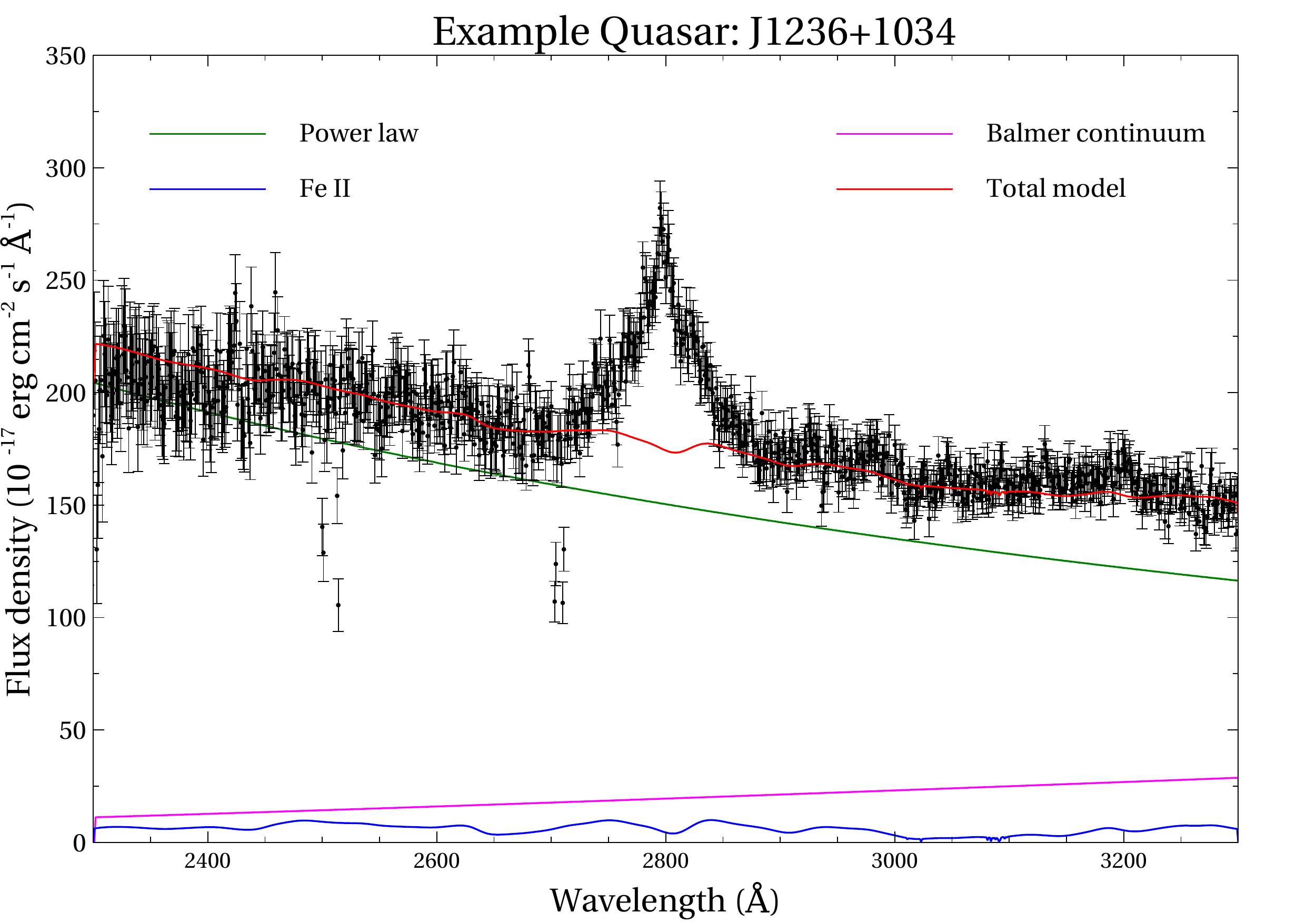}}
\caption{Example fitting of the SBB region for quasar SDSS J123604.51$+$103449.2 in the object rest-frame. Colored solid lines correspond to individual fitted continuous-emission components as indicated in the image, while the black circles (with errorbars) are the observed data points.}
\label{fig:sbbeg}
\end{figure}

\subsection{The Mg~{\sc ii} line}

Once the underlying SBB emission had been accurately reconstructed in each quasar spectrum, the Mg~{\sc ii} ${\rm \lambda{}}$2798 line could be modelled using Gaussian components. After subtraction of the disk power-law and total SBB emission, the remaining Mg~{\sc ii} emission was fitted using chisquare minimization. All objects required one or two Gaussians to achieve a good fit, with all components used being broad ($>$1000~km~s$^{-1}$) as no significant improvement to the fits was achieved by including narrow components. We assign no physical meaning to the individual properties of each component in cases where two Gaussians were required, they are simply used to construct the total profile. Using this profile, the FWHM and the wavelength span of the line could easily be identified. An example of a two-component Mg~{\sc ii} line fit is shown in Fig.~\ref{fig:mg2plot}.

\begin{figure}
\centering
\resizebox{\hsize}{!}{\includegraphics[width=8cm]{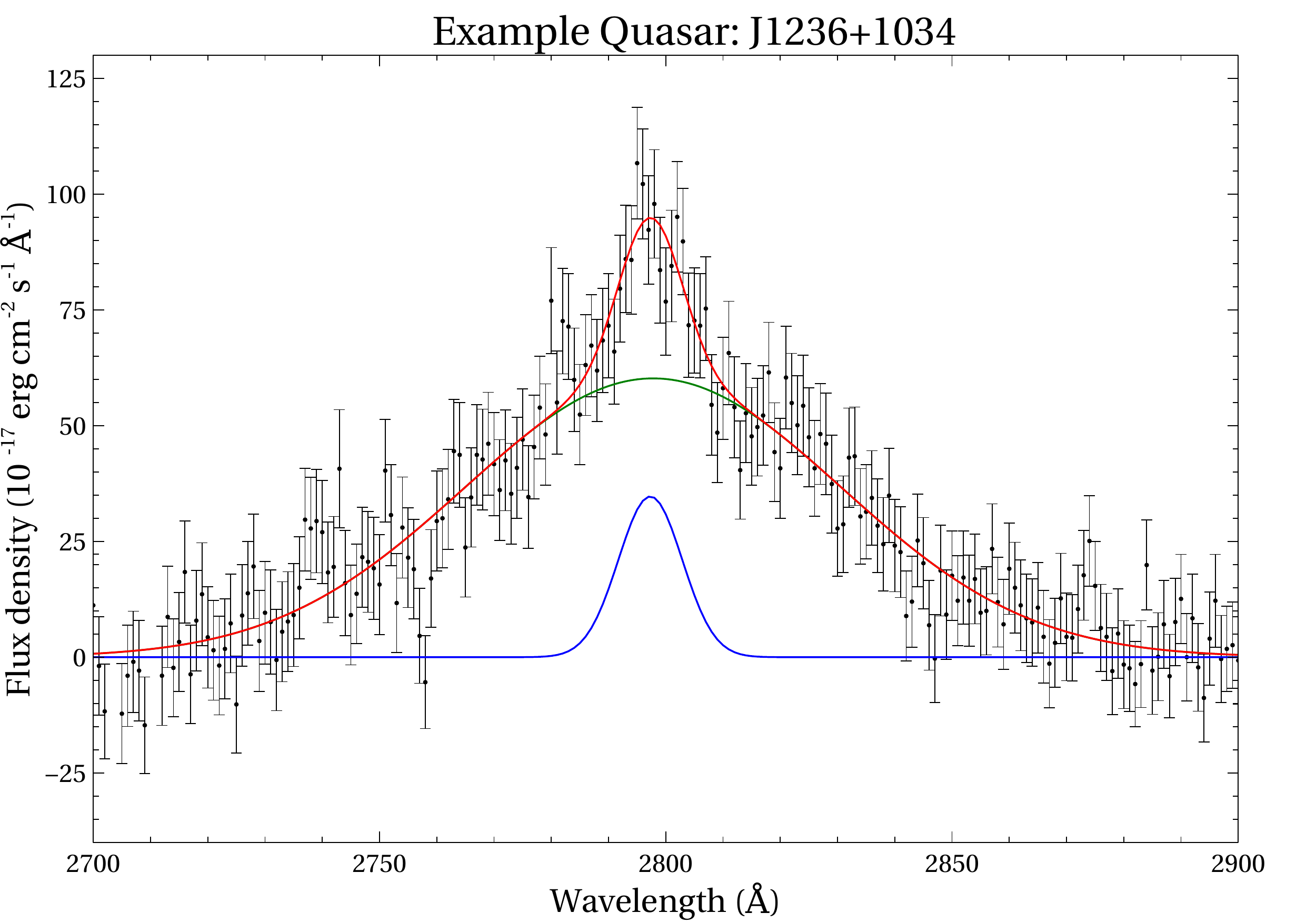}}
\caption{Example of Mg~{\sc ii} line fitting for quasar SDSS J123604.51$+$103449.2 in the object rest-frame. Black circles (with errorbars) indicate the observed data points after subtraction of the continuous-emission components (blended Fe~{\sc ii}, Balmer continuum and power-law). The green and blue solid lines indicate individual Gaussian components, while the red solid line is the sum of these components, giving the total profile.}
\label{fig:mg2plot}
\end{figure}

\subsection{The [O~{\sc iii}] doublet and H${\rm \beta}$ blend}

Optical [O~{\sc iii}] emission in quasars appears as doublet components located at ${\rm \lambda{}}$4959~\AA{} and ${\rm \lambda{}}$5007~\AA{} and is a very useful feature to analyze given its role in the EV1 correlation and its potential use as an orientation indicator. However, the redshift range spanned by our sample means only five of our objects have adequate SDSS spectral coverage of the doublet. In those five objects we fitted the [O~{\sc iii}] emission, together with the blended broad H${\rm \beta{}}$ emission, using one or two Gaussians for each of the [O~{\sc iii}] components and one Gaussian each for the broad and narrow components of H${\rm \beta{}}$, after subtraction of the power-law continuum. No Fe~{\sc ii} emission was obvious in the blend, so it was not included. The wings of the H${\rm \beta}$ line extend smoothly onto the power-law fitted in the UV part of the spectrum, indicating starlight is not significant in the wavelength range spanned by the H${\rm \beta{}}$ blend, hence it was not included. Cases where two Gaussian profiles were needed to replicate each of the [O~{\sc iii}] components follow the established pattern of a broader, blueshifted component blended with a narrow component whose centroid falls close to the systematic redshift \citep{bian05}. When using chisquare minimization to fit the total H${\rm \beta{}}$ blend, the narrow (and where applicable, broad) Gaussians of each [O~{\sc iii}] doublet component had their respective widths locked across the two components, and the flux of each Gaussian contributing to the red component was set to 3 times that of the corresponding Gaussian in the blue, as atomic theory predicts. The final fitted model profiles for each of the five objects are shown in Fig~\ref{fig:o3plot}.

\begin{figure}
\centering
\resizebox{\hsize}{!}{\includegraphics[width=8cm]{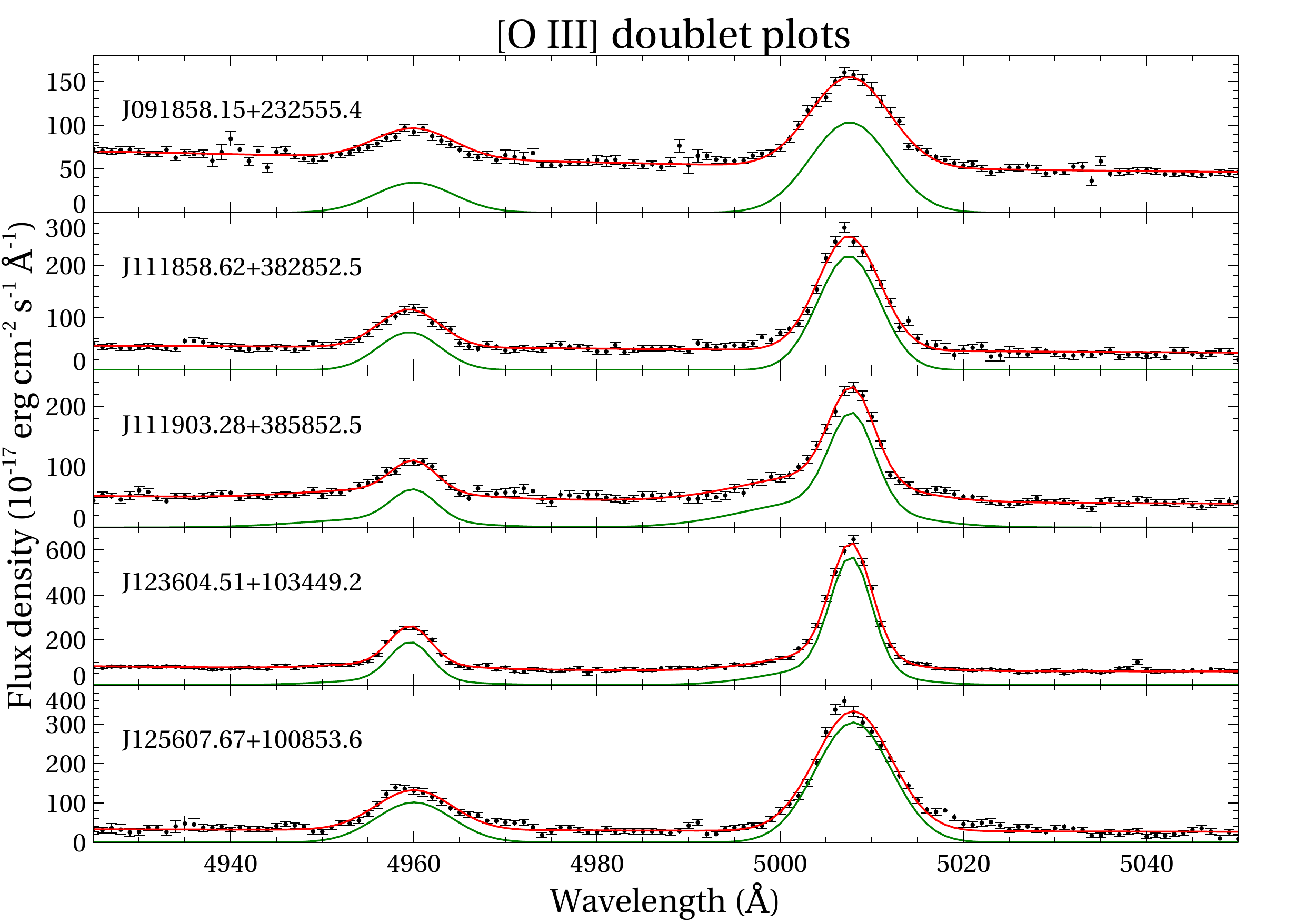}}
\caption{Plots of the [O~{\sc iii}] doublet fitting regions for all five quasars (vertically descending) which have adequate spectral coverage of this region. Quasar name is indicated in the top left of each plot Black circles (with errorbars) indicate the observed data points, while the red line indicates the total spectral model profile. The green line indicates the [O~{\sc iii}] doublet model.}
\label{fig:o3plot}
\end{figure}

\section{Analysis}

\subsection{Line properties and correlation testing}

We examine several line properties, as well as quasar inclination angle, in order to test for correlations. For the Mg~{\sc ii} ${\rm \lambda{}}$2798 line these properties are: FWHM, line dispersion (${\rm \sigma{}_{line}}$), FWHM-to-${\rm \sigma{}_{line}}$ ratio (FWHM/${\rm \sigma{}_{line}}$) and EW. In addition, the Fe~{\sc ii} EW and the ratio EW(Fe~{\sc ii})/EW(Mg~{\sc ii}) are also tested, giving a total of 21 property pairs. The EW of Mg~{\sc ii} is calculated over the wavelength range where the line model flux is at least 1 per cent of its peak flux, while the Fe~{\sc ii} EW is calculated over the entire SBB and Mg~{\sc ii} fitting region (2300--3300 \AA{}). Both FWHM and ${\rm \sigma{}_{line}}$ are methods of measuring the line width, with FWHM being sensitive to the line core, while ${\rm \sigma{}_{line}}$ is comparatively more sensitive to the line wings. Therefore, the FWHM-to-${\rm \sigma{}_{line}}$ ratio is an indicator of the line shape. Values of the FWHM-to-${\rm \sigma{}_{line}}$ ratio are well known for simple idealized line profiles, and become larger as the profile shape becomes more core dominated and vice-versa. For example, a perfectly rectangular profile has FWHM/${\rm \sigma{}_{line}}$=3.46, while a Gaussian has FWHM/${\rm \sigma{}_{line}}$=2.35.

Correlation tests between each pair of properties are performed using \emph{Spearman's rank correlation coefficient}. Where a significant correlation is found, a positive coefficient value indicates a positive correlation, while a negative value indicates a negative correlation. Calculation of these coefficients requires a unique value of each variable being tested, however, as can be seen in Table~\ref{tab:quasars}, some quasars in our sample share the same inclination angle. Where this occurs, all other properties undergoing correlation testing have their values averaged over all quasars sharing a particular inclination angle, before testing for correlation of these properties with inclination angle. Table~\ref{tab:cmatrix} shows the correlation coefficient for each pair of variables and indicates where this value corresponds to p(null), the probability of no correlation, being less than 5 per cent. For 21 uncorrelated pairs, one would expect to find one value satisfying this condition by chance, however this occurs six times for our sample, indicating that these pairs are worthy of further investigation. 

\begin{table*}
\begin{center}
\caption{Matrix of Spearman's rank correlation coefficients calculated for various line properties as well as inclination angle.}
\begin{tabular}{l l l l l l l l}
\hline\hline &Incl. angle&Mg~{\sc ii} FWHM&Mg~{\sc ii} ${\rm \sigma{}_{line}}$&Mg~{\sc ii} $\frac{\rm FWHM}{\rm \sigma{}_{line}}$&EW(Mg~{\sc ii})&EW(Fe~{\sc ii})&$\frac{\rm EW(Fe~{\rm II})}{\rm EW(Mg~{\rm II})}$\\
\hline
incl. angle& &$-$0.018&\phantom{$-$}0.015&\phantom{$-$}0.052&\phantom{$-$}0.293&\phantom{$-$}0.267&\phantom{$-$}0.076\\ \\
Mg~{\sc ii} FWHM&$-$0.018&&\phantom{$-$}0.313$^{\dagger{}a}$&\phantom{$-$}0.550$^{\dagger{}b}$&\phantom{$-$}0.127&$-$0.189&$-$0.189\\ \\
Mg~{\sc ii} ${\rm \sigma{}_{line}}$&\phantom{$-$}0.015&\phantom{$-$}0.313$^{\dagger{}a}$& &$-$0.565$^{\dagger{}c}$&\phantom{$-$}0.122&$-$0.246&$-$0.242\\ \\
Mg~{\sc ii} $\frac{\rm FWHM}{\rm \sigma{}_{line}}$&\phantom{$-$}0.052&\phantom{$-$}0.550$^{\dagger{}b}$&$-$0.565$^{\dagger{}c}$& &$-$0.055&$-$0.019&\phantom{$-$}0.028\\ \\
EW(Mg~{\sc ii})&\phantom{$-$}0.293&\phantom{$-$}0.127&\phantom{$-$}0.122&$-$0.055& &\phantom{$-$}0.393$^{\dagger{}d}$&$-$0.395$^{\dagger{}e}$\\ \\
EW(Fe~{\sc ii})&\phantom{$-$}0.267&$-$0.189&$-$0.246&$-$0.019&\phantom{$-$}0.393$^{\dagger{}d}$& &\phantom{$-$}0.636$^{\dagger{}f}$\\ \\
$\frac{\rm EW(Fe~{\rm II})}{\rm EW(Mg~{\rm II})}$&\phantom{$-$}0.076&$-$0.189&$-$0.242&\phantom{$-$}0.028&$-$0.395$^{\dagger{}e}$&\phantom{$-$}0.636$^{\dagger{}f}$& \\
\hline    
\end{tabular}
\label{tab:cmatrix}
\end{center}
$^{\dagger{}x}$ Condition (p(null)$\leq$0.05) is met, where $x$ is one of $a$, $b$, $c$, $d$, $e$, $f$ and identifies individual correlated pairs 
\end{table*}

The fact that 11 of the 56 quasars eligible for analysis were rejected warrants extra testing to ensure, as far as possible, that their exclusion is not due to factors which significantly influence the measured properties in this study. Such an outcome could introduce biases into the correlation results listed in Table~\ref{tab:cmatrix}. The catalog of \citet{shen11} contains data on over 100\,000 SDSS quasars, including all 11 excluded objects and 42 out of the 45 sample objects. Three relevant variables are listed for these quasars, namely Mg~{\sc ii} FWHM and both Fe~{\sc ii} and Mg~{\sc ii} equivalent width. Their method used to calculate these values is not identical to ours, however they do provide a way of testing for evidence of bias by providing values across both tested and excluded quasars. A Kolmogorov-Smirnov test indicates that both the excluded and sample quasars are consistent with having been drawn from the same distribution for each of the three measurement categories obtained from \citet{shen11} as well as the inclination angle. Additionally, when combined to form a sample containing both included and excluded quasars, significant correlations are only found in instances already noted from the original analysis sample data. Therefore there is no evidence that removing the 11 objects in question affected the correlation findings.

The criteria for significant correlation is not met in any case where inclination angle is tested against another measured property. It is a somewhat surprising result that no correlation is found between Mg~{\sc ii} FWHM and inclination angle due to the numerous studies which claim line width dependence on the apparent orientation of quasars. This lack of correlation is also evident upon examination of the FWHM values calculated in Paper~I and warrants further discussion (see Section~5.2). The positive relationship between the equivalent widths of Mg~{\sc ii} and Fe~{\sc ii} is easily explained, since both ions require photons of very similar energies to enable their creation (7.9~eV for Fe~{\sc i}$\rightarrow$Fe~{\sc ii} and 7.6~eV for Mg~{\sc i}$\rightarrow$Mg~{\sc ii}) and destruction (16.2~eV for Fe~{\sc ii}$\rightarrow$Fe~{\sc iii} and 15.0~eV for Mg~{\sc ii}$\rightarrow$Mg~{\sc iii}). Differences in the ionization environment between objects will therefore likely change the observed strengths of both ions' equivalent widths in the same sense (both strengthening or weakening). Unsurprisingly, as they are both methods of measuring the line width, there is apparent a positive correlation between FWHM and ${\rm \sigma{}_{line}}$.

The two types of line width measurement used in this study (FWHM and line dispersion) were examined by \citet{collin06} (hereafter C06) using the sample of H${\rm \beta{}}$-reverberation-mapped quasars listed in \citet{peterson04}. By doing this they attempted to determine a reliable method for calculating black hole mass using the properties of broad emission lines. Their study revealed that both measures of line width were found to vary between repeat observations of the same object, indicating that both values cannot depend solely on black hole mass, inclination or a combination of both. The same was also found to be true for the FWHM-to-${\rm \sigma{}_{line}}$ ratio. As is the case for our observations of Mg~{\sc ii} lines, a positive correlation between FWHM and ${\rm \sigma{}_{line}}$ was evident for H${\rm \beta{}}$ in C06. There are also correlations involving FWHM/${\rm \sigma{}_{line}}$ with respect to both FWHM and ${\rm \sigma{}_{line}}$. However, the manifestation of the relationship between FWHM-to-${\rm \sigma{}_{line}}$ ratio and line dispersion is very different from that for H${\rm \beta{}}$ in C06, which showed a positive correlation between the two variables (see fig.~3 of that paper), whereas in our case the correlation is negative. It should be noted, however, that the ${\rm \sigma{}_{line}}$ values in our Mg~{\sc ii} sample extend to much greater velocities than for H${\rm \beta{}}$ in C06, while the number of larger-than-Gaussian FWHM-to-${\rm \sigma{}_{line}}$ ratios recorded is small. 

In C06 objects are divided into categories for each of the line width measurements, for FWHM-to-${\rm \sigma{}_{line}}$ ratio they are Population 1 (FWHM/${\rm \sigma{}_{line}}$$<$2.35) and Population 2 (FWHM/${\rm \sigma{}_{line}}$$>$2.35), while for line dispersion they are Population A (${\rm \sigma{}_{line}}$$<$2000) and Population B (${\rm \sigma{}_{line}}$$>$2000). Most objects in C06 lay in either both Population 1 and Population A or both Population 2 and Population B. Conversely, based on the categorization of C06, 39 out of our 45 objects have the Mg~{\sc ii} line residing in the classes of both Population 1 and Population B. The greater width in our sample is likely due to the \citet{peterson04} quasars used in C06 being selected for low redshift to allow optical observation of the H${\rm \beta{}}$ line. Unlike the case for line dispersion, there is a positive correlation between FWHM and the FWHM-to-${\rm \sigma{}_{line}}$ ratio.

Both of the line width correlations with the FWHM-to-${\rm \sigma{}_{line}}$ ratio have very strong significance, as the probability for no correlation is p=0.0001 in both cases. This likely results from the obvious fact that the ratio's numerator is the FWHM, with which the correlation is positive, and the denominator is the line dispersion, with which the correlation is negative, together with the substantial scatter in the FWHM vs. ${\rm \sigma{}_{line}}$ relationship. The significance of the relationship between these two line width measurements is substantially smaller than the significance of the ratio's correlations with either of the line width measurements, a no-correlation probability being p=0.04. Therefore, the finding of positive correlation with FWHM and negative correlation with ${\rm \sigma{}_{line}}$ probably does not tell us anything given that the ratio measures FWHM divided by ${\rm \sigma{}_{line}}$.

The ratio of Fe~{\sc ii}-to-Mg~{\sc ii} EWs has been investigated previously for its suitability as an indicator of the Fe-to-${\rm \alpha{}}$ element abundance ratio \citep{verner03}. Observations could then be used for comparison with models which suggest timescales for Fe enrichment are longer than those for ${\rm \alpha{}}$-elements including Mg \citep{hamann93}. However, there appears to be no relationship between the ratio and redshift \citep{freudling03}. Our results give two significant correlations, one with each of the EW terms already in the ratio, namely those of Fe~{\sc ii}, where there is positive trend, and Mg~{\sc ii}, where there is a negative trend. Since Fe~{\sc ii} is the numerator and Mg~{\sc ii} is the denominator in the ratio, these correlations may not be telling us much other than the obvious fact that the EWs of Fe~{\sc ii} and Mg~{\sc ii} are not perfectly correlated, similar to the opposite-direction correlations found for the FWHM-to-${\rm \sigma{}_{line}}$ ratio with each measure of line width. In summary, it is possible to pick out only two significant correlations found in this section which have physical significance. They are: (i) the positive correlation between EW(Mg~{\sc ii}) and EW(Fe~{\sc ii}), and (ii) the positive correlation between Mg~{\sc ii} FWHM and the line dispersion. These are illustrated in Fig.~\ref{fig:corrs}.

\begin{figure}
\centering
\resizebox{\hsize}{!}{\includegraphics[width=8cm]{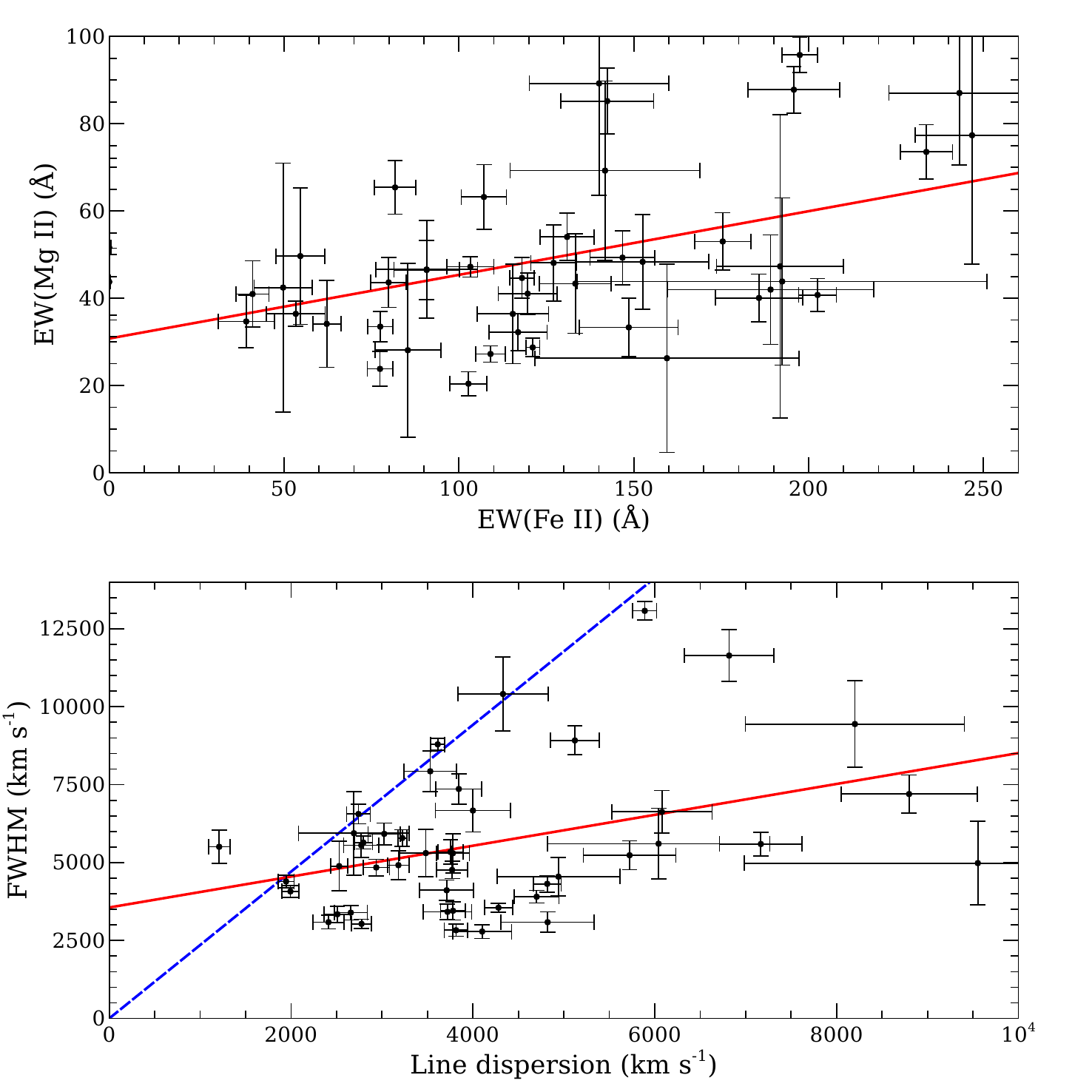}}
\caption{Plots showing property pairs which meet the criteria for significant correlation. \emph{Top panel}: relationship between equivalent widths of Mg~{\sc ii} and Fe~{\sc ii}. \emph{Bottom panel}: relationship between Mg~{\sc ii} FWHM and line dispersion. Best fit linear trendlines are in red. The blue dashed line indicates allowed parameters for a Gaussian profile.}
\label{fig:corrs}
\end{figure}

\subsection{Broad lines at high inclination}

In common AGN unification schemes, the visibility of unpolarized broad lines indicates a low angle between the accretion disk axis and the line-of-sight of the observer. At high angles the line-of-sight to the central engine is then obscured by a dusty torus lying in the plane of the accretion disk and is only visible in polarized light, as seen in NGC~1068 \citep{antonucci85}. Our sample includes many objects which have an inclination angle greater than that predicted to be possible for Type~1 objects, for example \citet{marin14} predicts a maximum of $\sim$60$^\circ{}$ beyond which obscuration of the disk and BLR occurs. This issue was also noted in Paper~I. Indeed, broad lines are visible in our sample objects at angles all the way up to those indicating an almost edge on view (89$^\circ{}$), as shown in Fig.~\ref{fig:widsvang}. As stated previously, there is no correlation between either line width measure and inclination angle.

\begin{figure}
\centering
\resizebox{\hsize}{!}{\includegraphics[width=8cm]{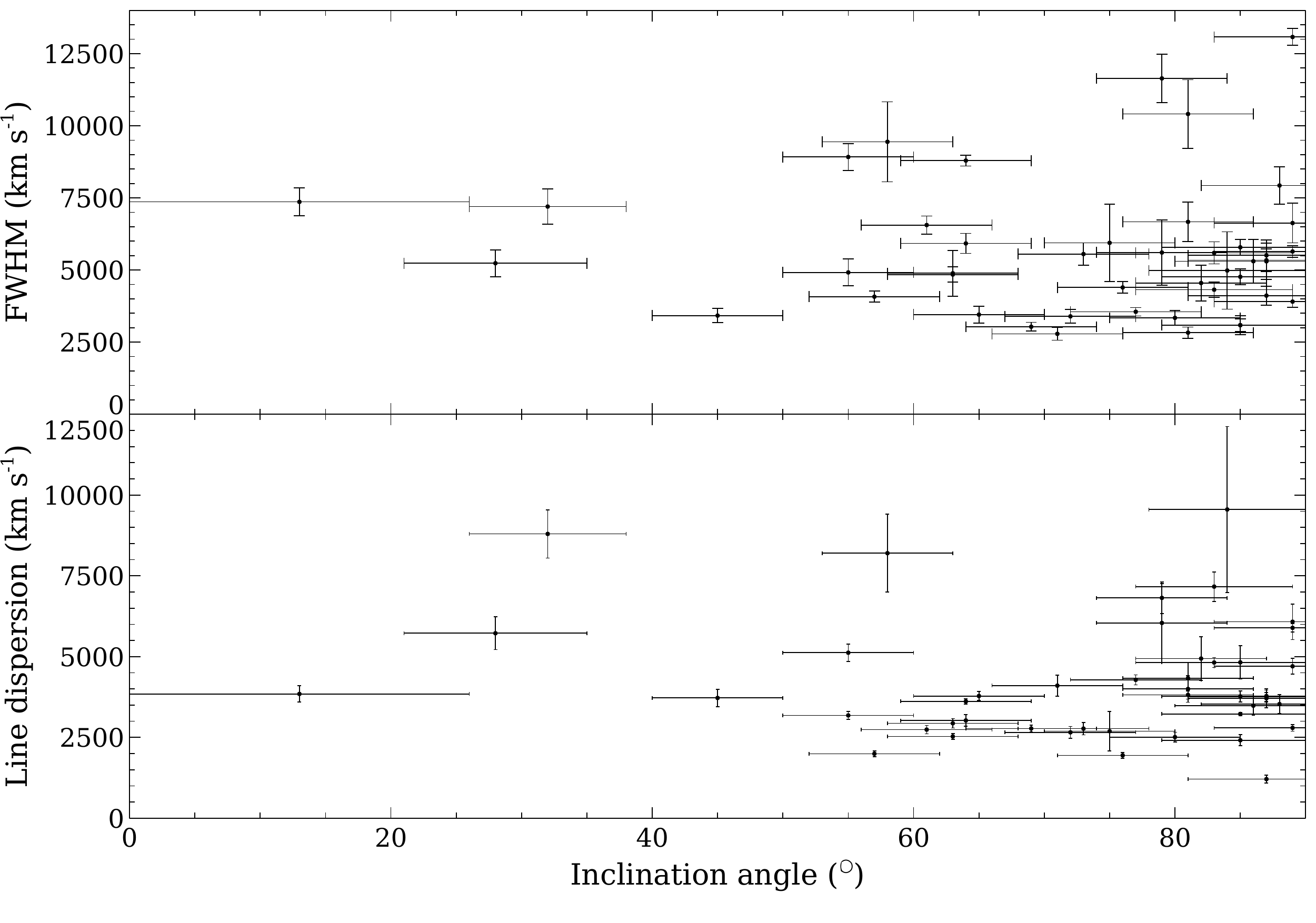}}
\caption{Plots of FWHM against inclination angle (top panel) and line dispersion against inclination angle (bottom panel) for Mg~{\sc ii} lines in each quasar. There is no maximum angle beyond which broad lines are not detected.}
\label{fig:widsvang}
\end{figure}

\section{Discussion}

\subsection{The [O~{\sc iii}] emission}

The results of \citet{risaliti11} and \citet{bisogni17} suggest [O~{\sc iii}] EW scales positively with quasar inclination angle. It was possible to calculate the equivalent width of the [O~{\sc iii}] ${\rm \lambda{}}$5007 doublet component for each of the five objects having spectral coverage of the line. The results are compared against inclination angle for each of the five objects as shown in Table~\ref{tab:o3ews}.

\begin{table}
\begin{center}
\caption{List of [O~{\sc iii}] equivalent width values}
\begin{tabular}{l l l}
\hline Object name&Inclination angle&[O~{\sc iii}] EW\\
 &($^{\circ}$)&(\AA{})\\
\hline
J091858.15$+$232555.4&81&24.8$\pm$2.8\\
J111858.62$+$382852.2&55&59.2$\pm$16.9\\
J111903.28$+$385852.5&64&46.6$\pm$8.3\\
J123604.51$+$103449.2&63&66.1$\pm$7.3\\
J125607.67$+$100853.6&69&123.6$\pm$17.1\\
\hline
\end{tabular}
\label{tab:o3ews}
\end{center}
\end{table}

It must be stressed that this is a very small number of objects and therefore trends cannot be identified involving inclination angle or any other variable examined in this paper. However, it should be noted that the [O~{\sc iii}] EWs recorded here are all large relative to the peak of the distribution recorded in \citet{risaliti11}, which occurs at $\sim$10~\AA{} (see fig.~1 in that paper). This supports the hypothesis that quasars viewed at comparatively edge-on orientations have large [O~{\sc iii}] EW, since these objects have relatively large inclination angles compared to the general observed quasar population. It also appears that the Fe~{\sc ii} emission within the H${\rm \beta{}}$ blend is comparatively weak, since adding Fe~{\sc ii} emission lines to the model is not necessary to achieve a good fit. Given these large [O~{\sc iii}] EW values, this supports the general anti-correlation between the strengths of [O~{\sc iii}] and Fe~{\sc ii} revealed by EV1 in \citet{boroson92}.

\subsection{What does the behavior of Mg~{\sc ii} line properties with respect to inclination angle tell us?}

A striking result from our investigation is that both measures of line width are consistent with having no correlation with inclination at angles all the way up to an edge-on viewing orientation. In fact, the correlation tests reveal the probability of no dependence to be p$>$0.9 in both cases. Aside from the questions posed regarding the validity of the unified model in the context of observable broad lines at high inclinations, the lack of correlation places restrictions on the possible geometric configuration of the region producing broad Mg~{\sc ii}. If the Mg~{\sc ii} emitting gas is located in a structure similar to a disk, then one would expect the projection of a BLR cloud's rotational velocity along the line-of-sight to depend on inclination angle according to $v_{obs}=v_{rot}sin\theta{}_{i}$, where $v_{obs}$ is the maximum observed velocity and $v_{rot}$ is the intrinsic rotational velocity. Therefore in the disk model the highest inclination objects should, in general, produce the largest line widths. The lack of correlation could mean that Mg~{\sc ii} FWHM is a good indicator of black-hole mass in lobe-dominated quasars, however a large Mg~{\sc ii} reverberation campaign would be required to confirm this hypothesis.

In \citet{shen14} the EV1 anti-correlation between the Fe~{\sc ii}/H${\rm \beta{}}$ strength ratio and [O~{\sc iii}] strength is interpreted to result from an increase in Eddington ratio in the direction of weakening [O~{\sc iii}], with a substantial spread in H${\rm \beta{}}$ FWHM observed at each point along the correlation axis (See fig.~1 in that paper). This FWHM spread is in turn portrayed as an orientation effect, with higher inclination angles resulting in greater line widths at fixed black hole mass. Such a view depends on H${\rm \beta{}}$ velocities projecting strongly into the plane perpendicular to the radio axis. The fact that there is no correlation with FWHM in our Mg~{\sc ii} data suggests this may not be the case in lobe-dominated quasars exhibiting broad UV/optical emission lines. An alternative possibility is that Mg~{\sc ii} occupies a less ``disk-like'' geometry than H${\rm \beta{}}$. How plausible is this? There are studies that suggest the Mg~{\sc ii} and H${\rm \beta{}}$ emission regions are not co-spatial, for example \citet{wang09} show that the Mg~{\sc ii} FWHM in a large sample of Seyfert~1s and quasars is typically significantly lower than that for H${\rm \beta{}}$, suggesting it forms at greater distances assuming a virialized gas. There is also evidence that the Mg~{\sc ii} line center is less shifted from the object redshift than H${\rm \beta{}}$ \citep{marziani13}. However, this argument is easily countered by the findings of previous studies such as \citet{ghisellini93} and \citet{aars05} regarding the dependence of Mg~{\sc ii} line width on inclination.

Although Mg~{\sc ii} EW as a function of inclination angle in our sample does not meet the criterion for significant correlation, the actual presence of such a correlation cannot be ruled out with high confidence. This is due to the probability of no correlation being rather low (p=0.12). If such a trend does in fact exist, then it is in the positive sense as depicted in Fig.~\ref{fig:ewmgvang}, similar to the findings for [O~{\sc iii}] in \citet{risaliti11}. This would imply that the Mg~{\sc ii} emission region radiates relatively isotropically compared to the accretion disk, causing the power-law continuum to appear weaker relative to Mg~{\sc ii} as inclination angle increases. Such a scenario is supported by the lack of correlation between line width and inclination angle, in contrast to findings for the general quasar population in previous studies. It is also notable that for Fe~{\sc ii} EW, the no-correlation probability with inclination angle is also not very high (p=0.16). The typical geometry of the low ionization BLR in our sample could therefore differ from the flat-disk configuration suggested for the general quasar population, for example by being approximately spherical or forming a highly distorted or thickened disk. The average Mg~{\sc ii} EW in our sample measures 48.8~\AA{}, which is similar to the value of 43.9~\AA{} found in a sample of 635 quasars selected from the Large Bright Quasar Survey reported in \citet{forster01}. This indicates that there is nothing unusual about the Mg~{\sc ii} line strengths in our sample compared to quasars overall. This seems to rule out a perfectly spherical configuration for the Mg~{\sc ii} emission region in our sample, as such a situation would generate atypically large EWs in cases of high inclination.

\begin{figure}
\centering
\resizebox{\hsize}{!}{\includegraphics[width=8cm]{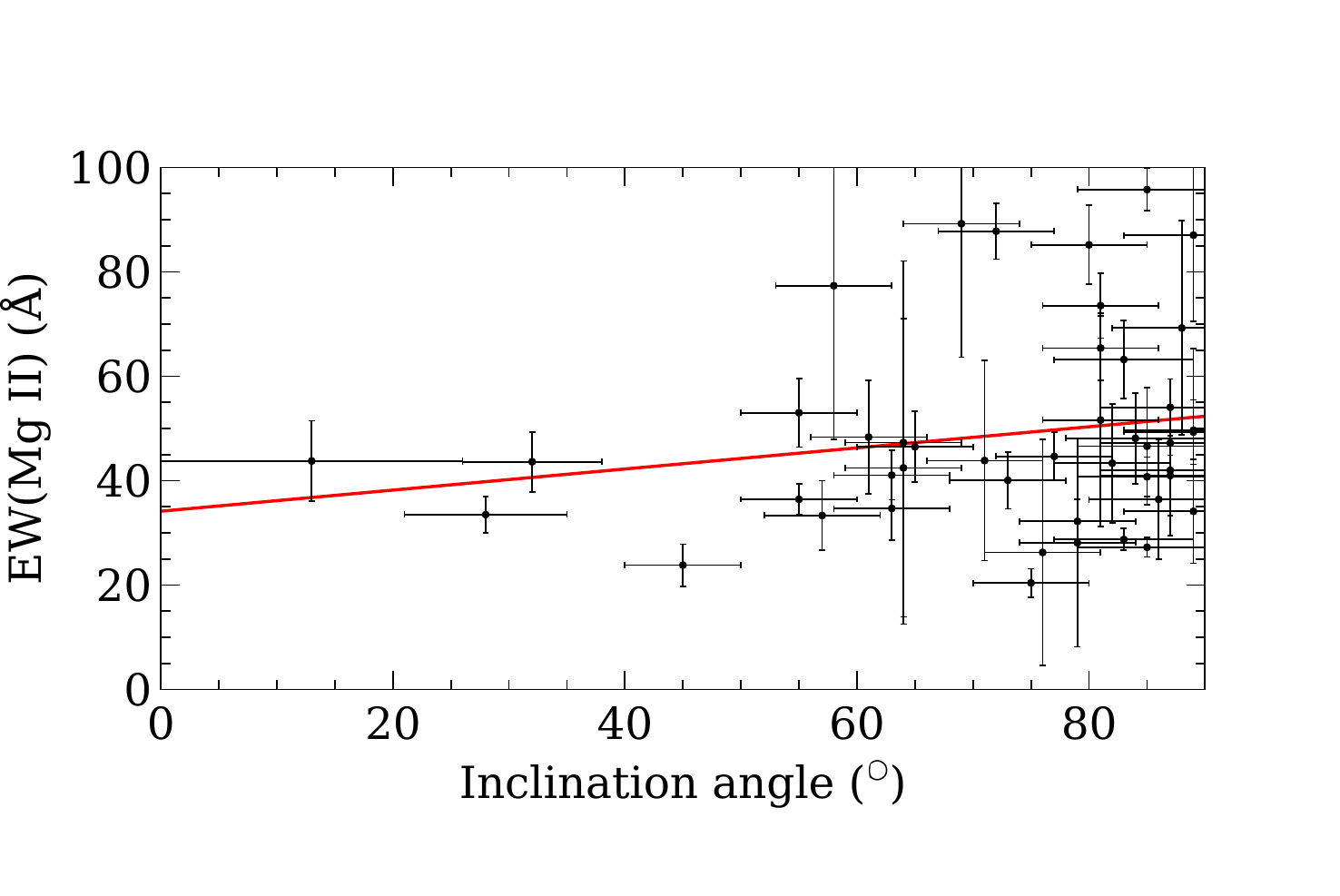}}
\caption{Plot of Mg~{\sc ii} EW against inclination angle for all quasars, with the best-fit linear trendline in red.}
\label{fig:ewmgvang}
\end{figure}

\subsection{Absence of an obscuring torus at high inclination}

The fact that broad emission lines can be seen at large inclination angles in our sample, for example Fig.~\ref{fig:quasar89}, contradicts the existence in these objects of a smooth, continuous dusty obscuring torus orientated co-planar with the accretion disk. In order to test for the presence of dust in our sample objects, mid-infrared data derived from the All-Sky survey undertaken by the Wide-field Infrared Survey Explorer (WISE) satellite \citep{wright10} was compared to each quasar's disk continuum by extending the optical power-law to 2~${\rm \mu{}}$m in the rest-frame. After correction for redshift, linear interpolation or extrapolation of the spectral flux recorded by WISE at the central wavelengths of the two shortest wavelength bands (3.4~${\rm \mu{}}$m and 4.6~${\rm \mu{}}$m) was used to find the total spectral flux at 2~${\rm \mu{}}$m. The 2~${\rm \mu{}}$m total flux to disk ratio could then be calculated. No correlation was found between this ratio and quasar inclination angle, the average ratio over all sample objects being 2.7. This is comparable to a rough estimate of $\sim$3 made by examining the mean spectral energy distribution corresponding to all SDSS quasars shown in fig.~11 of \citet{richards06} and extending a straight line from the red side of the big blue bump peak to 2~${\rm \mu{}}$m. This indicates that, although the sample quasars are unobscured at high inclination, their dust content is not unusual among the quasar population.

\begin{figure}
\centering
\resizebox{\hsize}{!}{\includegraphics[width=8cm]{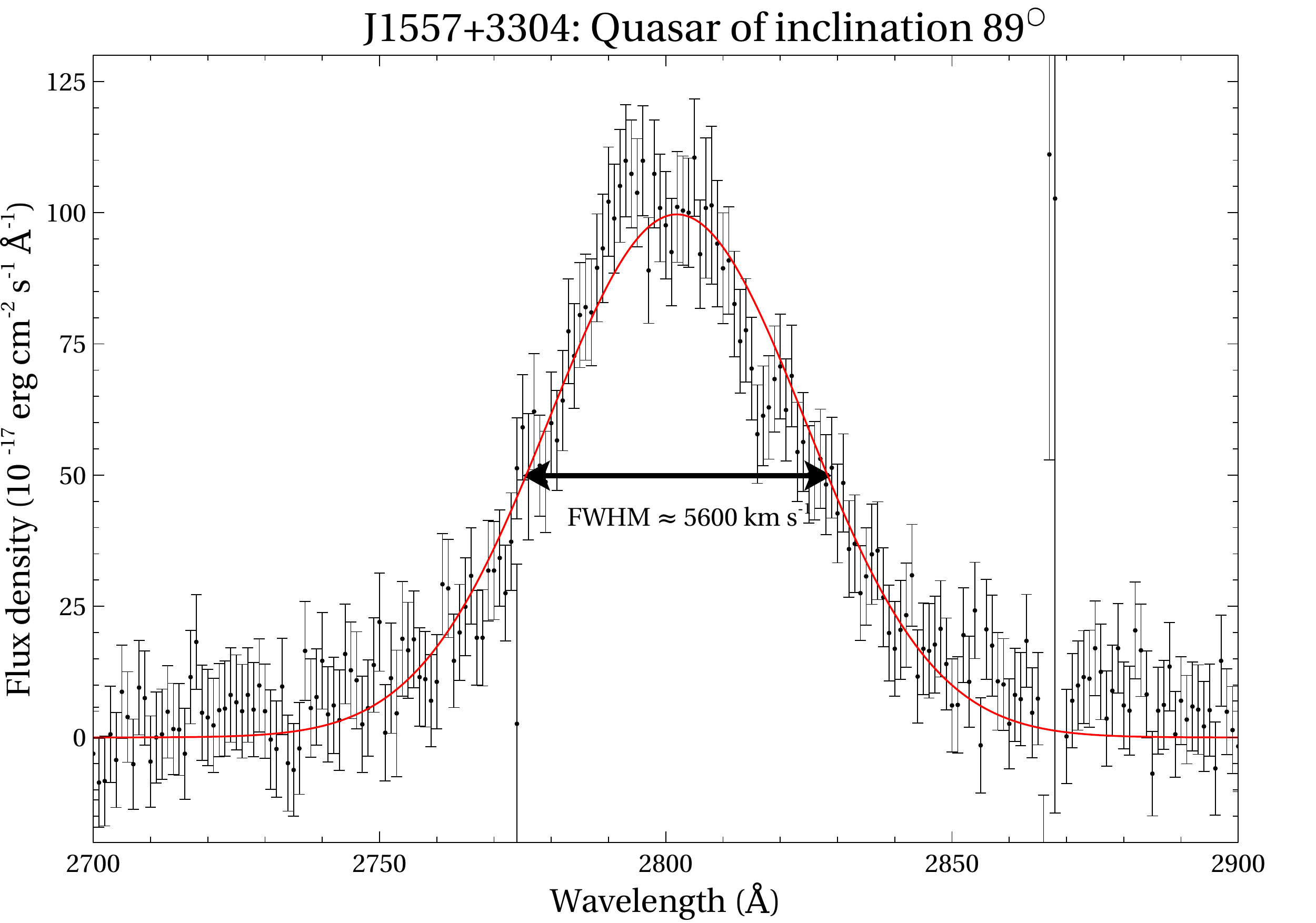}}
\caption{Continuous-emission-subtracted spectrum of quasar SDSS J155729.93$+$330446.9 (black circles with errorbars) which has an almost edge-on inclination angle of 89$^{\circ{}}$. Despite this, the broad Mg~{\sc ii~} line is visible. The solid red line is the Mg~{\sc ii} broad line model fit.}
\label{fig:quasar89}
\end{figure}

A possible explanation for these results could be a ``clumpy'' torus where increasing viewing angle increases the probability of obscuration of the central engine. Unlike the smooth torus model, this probability never becomes zero even when the viewing angle reaches edge-on \citep{elitzur08}. Clumpy torus models have become popular in recent years due to models of the gas inflow towards the central engine from the surrounding galactic environment \citep{netzer15}. The possibility of warped obscuring disks \citep{lawrence10} may also provide an explanation, since highly distorted dusty structures could disproportionately cover regions of the sky close to the radio axis as seen from the AGN in comparison to the traditional co-planar torus model. This could leave equatorial lines-of-sight to the central engine unobscured.

As noted in Section~2, lobe-dominated quasars of the type used in our sample are rare within the quasar population. It could therefore be that the vast majority of high inclination quasars are obscured by dust in the accretion disk equatorial plane. Searches for lobe-dominated objects would then record only those rare quasars which give an unobscured view of the central engine at high inclination. As noted previously in this study and in Paper~I ,the properties of our sample quasars, including their dust content, are not unusual apart from the lack of FWHM correlation with inclination and the high inclination angles themselves. Therefore it would seem that the simplest explanation is that our sample records those high inclination quasars which happen to have a dust geometry allowing observation of the central engine.

\section{Summary}

This section provides a brief summary of the main conclusions drawn in this paper:\\

\noindent (i) There is no correlation between the width of Mg~{\sc ii} ${\rm \lambda{}}$2798, as measured by either FWHM or line dispersion, and object inclination angle in our sample. This suggests that the low ionization BLR in lobe-dominated quasars is not confined to a thin disk geometry, unlike what has been found previously for the general quasar population.

\noindent (ii) The average Mg~{\sc ii} EW in our sample is not much higher than for quasars generally. Given the preponderance of high inclination objects recorded in this study, a spherical geometry for the Mg~{\sc ii} broad emission region would seem to be ruled out. It may instead occupy a heavily distorted or thickened disk.

\noindent (iii) Any correlation between Mg~{\sc ii} EW and inclination angle does not meet the strict criteria for significance. However it would be a positive correlation, and cannot be ruled out with any confidence. Therefore this finding does not provide strong evidence against the conclusions in (i).

\noindent (iv) The lack of correlation in (i) could indicate that Mg~{\sc ii} FWHM is a black hole mass indicator unaffected by orientation in lobe-dominated quasars. Further evidence from large future Mg~{\sc ii} reverberation mapping campaigns are needed to confirm this.

\noindent (v) The five [{O~\sc iii}] EW measurements reveal numbers which are significantly higher than that of the peak of the distribution recorded in the general quasar population. This supports the hypothesis that highly inclined quasars have high [O~{\sc iii}] EWs. It also concurs with the EV1 correlation, since Fe~{\sc ii} emission is weak in these objects.

\noindent (vi) All objects in our sample have unobscured lines-of-sight to the central engine, even those at edge-on inclinations. In such objects a smooth dusty torus co-planar with the accretion disk is insufficient to explain such occurrences, which may be instead accounted for by a clumpy torus or a warped disk obscurer, allowing sight-lines to the inner regions in at least some circumstances. This may be a selection effect resulting from the large inclination angles of our sample objects, allowing only those rare quasars having unobscured central engines at high viewing angle to be observed.

\section*{Acknowledgements}

We wish to thank the anonymous referee for their helpful comments which led to the improvement of this paper. This work was supported by the Polish Funding Agency National Science Center, project 2015/17/B/ST9/03436/ (OPUS 9) and is based on spectroscopic observations made by the Sloan Digital Sky Survey. Funding for the Sloan Digital Sky Survey IV has been provided by the Alfred P. Sloan Foundation, the U.S. Department of Energy Office of Science, and the Participating Institutions. SDSS-IV acknowledges support and resources from the Center for High-Performance Computing at the University of Utah. The SDSS web site is www.sdss.org. SDSS-IV is managed by the Astrophysical Research Consortium for the Participating Institutions of the SDSS Collaboration including the Brazilian Participation Group, the Carnegie Institution for Science, Carnegie Mellon University, the Chilean Participation Group, the French Participation Group, Harvard-Smithsonian Center for Astrophysics, Instituto de Astrof\'isica de Canarias, The Johns Hopkins University, Kavli Institute for the Physics and Mathematics of the Universe (IPMU) / University of Tokyo, Lawrence Berkeley National Laboratory, Leibniz Institut f\"ur Astrophysik Potsdam (AIP), Max-Planck-Institut f\"ur Astronomie (MPIA Heidelberg), Max-Planck-Institut f\"ur Astrophysik (MPA Garching), Max-Planck-Institut f\"ur Extraterrestrische Physik (MPE), National Astronomical Observatories of China, New Mexico State University, New York University, University of Notre Dame, Observat\'ario Nacional / MCTI, The Ohio State University, Pennsylvania State University, Shanghai Astronomical Observatory, United Kingdom Participation Group, Universidad Nacional Aut\'onoma de M\'exico, University of Arizona, University of Colorado Boulder, University of Oxford, University of Portsmouth, University of Utah, University of Virginia, University of Washington, University of Wisconsin, Vanderbilt University, and Yale University.

\software{IRAF \citep{iraf93}}

\bibliography{bib_cw}

\bibliographystyle{aastex61}


\end{document}